\documentclass[twocolumn,aps,pra,groupedaddress,amsmath,amssymb,showpacs]{revtex4-1}
\usepackage{graphicx}
\usepackage{dcolumn}
\usepackage{bm}
\usepackage{color}
\usepackage{multirow}
\usepackage[mathscr]{euscript}
\usepackage{subfigure}
\usepackage[colorlinks=true,allcolors=blue]{hyperref}
\usepackage{braket}

\newcommand{\hh}{\hat{H}}
\newcommand{\BE}{\begin{eqnarray}}
\newcommand{\EE}{\end{eqnarray}}
\newcommand{\ada}{\hat{a}_{\mathbf{k}}^{\dagger}\hat{a}_{\mathbf{k}}}
\newcommand{\dagg}{^{\dagger}}
\newcommand{\ek}{\epsilon_{k}}
\newcommand{\1}{&=&}
\newcommand{\2}{&+&}

\newcommand{\ee}[1]{\mathrm{e}^{#1}}
\newcommand{\ah}[1]{\hat{a}_{#1}}
\newcommand{\bh}[1]{\hat{b}_{#1}}

\newcommand{\tek}{\epsilon_{k}}

\newcommand{\tnc}{n_{\mathrm{c}}(t)}
\newcommand{\nn}{\nonumber}
\newcommand{\ii}{\text{i}}
\newcommand*\dif{\mathop{}\!\mathrm{d}}

\newcommand{\gt}{g^{(2)}}

\newcommand{\vtwo}{|v_{\mathbf{k}}|^2}

\newcommand{\tgt}{g^{(2)}}

\newcommand\rsout{\bgroup\markoverwith{\textcolor{red}{\rule[0.5ex]{4pt}{1pt}}}\ULon}

\newcommand{\nc}{n_{\mathrm c}}

\begin{document}
\title{Dynamical excitation of maxon and roton modes in a Rydberg-Dressed\\ Bose-Einstein Condensate}

\author{Gary McCormack$^{1}$, Rejish Nath$^{2}$ and Weibin Li$^{1}$} 
\affiliation{$^{1}$School of Physics and Astronomy, and Centre for the Mathematics and Theoretical Physics of Quantum Non-Equilibrium Systems, University of Nottingham, NG7 2RD, UK\\ $^2$Indian Institute of Science Education and Research, Pune, 411008, India}%

\begin{abstract}
	We investigate the dynamics of a three-dimensional Bose-Einstein condensate of ultracold atomic gases with a soft-core shape long-range interaction, which is induced by laser dressing the atoms to a highly excited Rydberg state. For a homogeneous condensate, the long-range interaction drastically alters the dispersion relation of the excitation, supporting both roton and maxon modes. Rotons are typically responsible for the creation of supersolids, while maxons are normally dynamically unstable in BECs with dipolar interactions. We show that maxon modes in the Rydberg-dressed condensate, on the contrary, are  dynamically stable. We find that the maxon modes can be excited through an interaction quench, i.e. turning on the soft-core interaction instantaneously. The emergence of the maxon modes is accompanied by oscillations at high frequencies in the quantum depletion, while rotons lead to much slower oscillations. The dynamically stable excitation of the roton and maxon modes leads to persistent oscillations in the quantum depletion. Through a self-consistent Bogoliubov approach, we identify the dependence of the maxon mode on the soft-core interaction. Our study shows that maxon and roton modes can be excited dynamically and simultaneously by quenching Rydberg-dressed long-range interactions. This is relevant to current studies in creating and probing exotic states of matter with ultracold atomic gases.
\end{abstract}

\maketitle

\section{Introduction}\label{sec:Introduction}
Collective excitations induced by particle-particle interactions play an important role in the understanding of static and dynamical properties of many-body systems. The ability to routinely create and precisely control properties of ultracold atomic gases opens exciting prospects to manipulate and probe collective excitations. In weakly interacting Bose-Einstein condensates (BECs) with s-wave interactions~\cite{Bolda2002,Jaksch2005,Pethick2008,PitaevskiiLev2016BCaS}, phonon excitations reduce the condensate density, giving rise to \textit{quantum depletion}~\cite{lee_1957}. It has been shown~\cite{Lopes2017} that quantum depletion can be enhanced by increasing the s-wave scattering length through Feshbach resonances~\cite{Stenger1999,Ketterle1998}. By dynamically changing the s-wave scattering length~\cite{Makotyn2014}, phonon excitations can alter the quantum depletion, the momentum distribution~
\cite{Martone2018}, correlations~\cite{Natu2013}, contact~\cite{Yin2013a,Sykes2014}, and statistics~\cite{Kain2014} of the condensate. Moreover the phonon induced quantum depletion plays a vital role in the formation of droplets in BECs~\cite{Cabrera301}.  

\begin{figure}
\centering
		\includegraphics[width=1.0\linewidth]{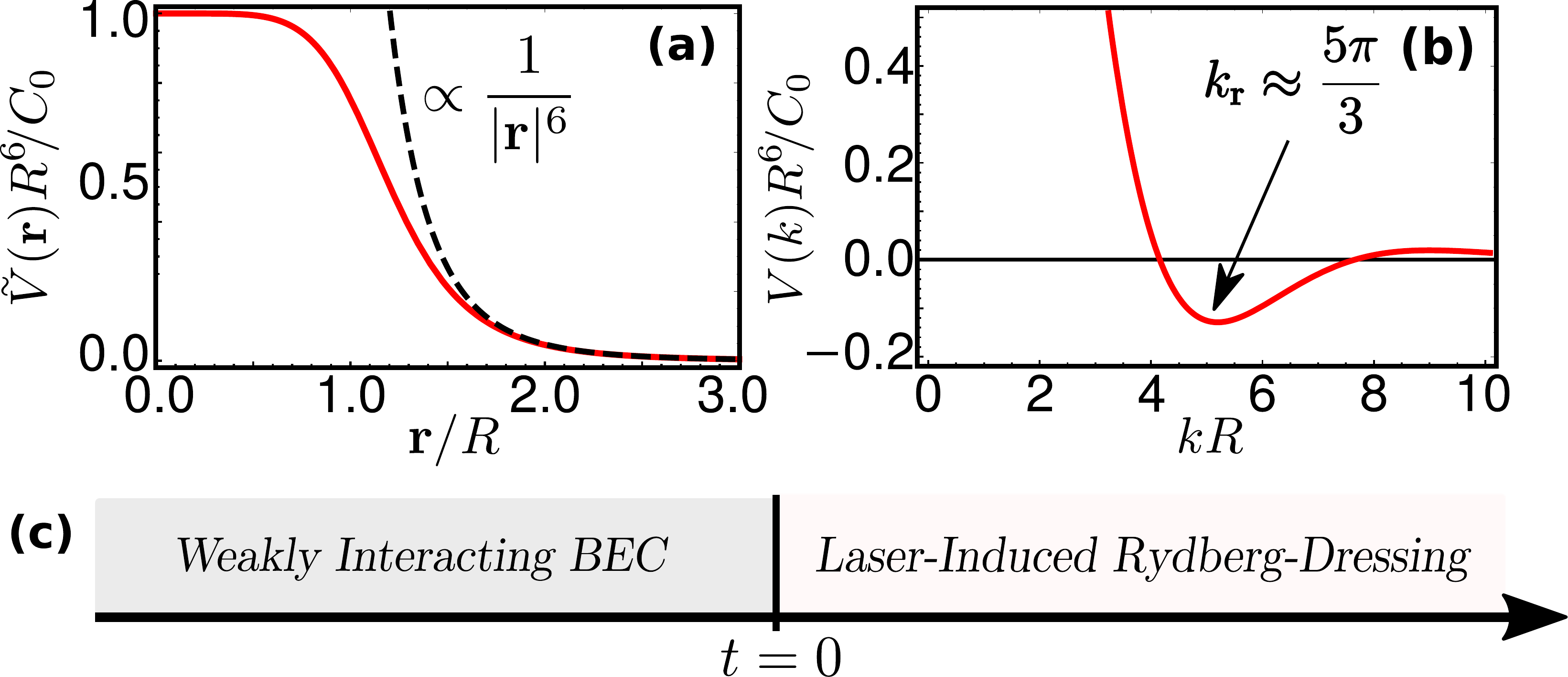}
			\caption{(color online) \textbf{Soft-core interaction and quench scheme}. (a) The soft-core interaction as a function of the interatomic distance $\bold{r}$. Energy is scaled by $R^6/C_0$ with $R$ and $C_0$ to be the soft-core radius and dispersion coefficient. The interaction is constant when $\bold{r}\ll R$, and becomes a vdW type when $\bold{r} \gg R$. (b) Fourier transformation of the soft-core interaction. The minimum of the interaction locates at $k_{\mathrm{r}} \approx 5\pi/3 R$, where the interaction is attractive. (c) The quench scheme. A weakly interacting BEC with s-wave interactions is first prepared. The laser dressing is applied at $g\le 0$, which induces the soft-core interaction. }\label{fig:softcore}
\end{figure}

When long-range interactions are introduced, the dispersion relation corresponding to the quasiparticle spectrum of a BEC is qualitatively different, where the excitation energies of the collective modes depend non-monotonically on the momentum. Previously BECs with dipole-dipole interactions have been extensively examined~\cite{Santos2000,Fischer2006a,Ticknor2011a,Lahaye2009,Natu2014a,Wilson2016,Cai2010}. In two-dimensional (2D)  dipolar BECs~\cite{santos_roton-maxon_2003}, {\it{roton}} and  {\it{maxon}} modes emerge, where roton (maxon) modes correspond to local minima (maxima) in the dispersion relation. The strength of dipolar interactions can be tuned by either external electric or magnetic fields~\cite{Lahaye2009}. When instabilities of roton modes are triggered, a homogeneous BEC undergoes density modulations such that a supersolid phase could form. The existence of roton modes has been supported by a recent experiment~\cite{Chomaz2018}.  Maxon modes, on the other hand, normally appear at lower momentum states~\cite{santos_roton-maxon_2003}. It was shown however that the maxon modes in dipolar BECs are typically unstable and decay rapidly through the Beliaev damping~\cite{Natu2014a,Wilson2016}.

Strong and long-range interactions are also found in gases of ultracold Rydberg atoms~\cite{Lesanovsky2011, Sela2011,  Weimer2010,li_nonadiabatic_2013,ates_dissipative_2012}. Rydberg atoms are in highly excited electronic states and interact via long-range van der Waals (vdW) interactions.  The strength of the vdW interaction is proportional to $\mathcal{N}^{11}$ with $\mathcal{N}$ to be the principal quantum number in the Rydberg state. For large $\mathcal{N}$ (current experiments exploit $\mathcal{N}$ typically between $30$ and $100$), the interaction between two Rydberg atoms can be as large as several MHz at a separation of several micrometers~\cite{saffman_quantum_2010}. However lifetimes in Rydberg states are typically $10\sim 100\mu$s, which is not long enough to explore spatial coherence. As a result, {\it{Rydberg-dressing}}, in which a far detuned laser couples electronic ground states to Rydberg states, is proposed. The laser coupling generates a long-range, {\it{soft-core}} type interaction between Rydberg-dressed atoms~\cite{Henkel2012,Maucher2011c,Honer2010,Cinti2010, Henkel2010a,Geisler2018, Lauer2012,Chougale2016, Gaul2016, Balewski2014}. The coherence time and interaction strength can be controlled by the dressing laser~\cite{Henkel2010a}. With this dressed interaction, interesting physics, such as magnets~\cite{Zeiher2016},  transport~\cite{Viteau2011}, supersolids~\cite{Pupillo2010,Henkel2012,Cinti2010,Li2018a}, etc, have been studied. Signatures of the dressed interaction have been experimentally demonstrated with atoms trapped in optical lattices and optical tweezers~\cite{Jau2016,Zeiher2016}. 

In this paper, we study excitations of roton and maxon modes in three dimensional (3D) Rydberg-dressed BECs in free space at zero temperature. Three dimensional uniform trapping potential of ultracold atoms have been realized experimentally~\cite{gaunt_bose-einstein_2013}. When the soft-core interaction is strong, both the roton and maxon modes are found in the dispersion relation of the collective excitations. Starting from a weakly interacting BEC, roton and maxon modes are dynamically excited by instantaneously switching on the Rydberg-dressed interaction. Through a self-consistent Bogoliubov calculation, we show that the roton and maxon modes leads to non-equilibrium dynamics, where the quantum depletion exhibits slow and fast oscillations. Through analyzing the Bogoliubov spectra, we identify that the slow oscillation corresponds to the excitation of the roton modes, while the fast oscillation comes from the excitation of the maxon modes. The dependence these modes have on the  quantum depletion in the long time limit is determined analytically and numerically.

The paper is organized as follows. In Sec.~\ref{sec:Hamiltonian}, the Hamiltonian of the system and  properties of the soft-core interaction are introduced. Bogoliubov methods, that are capable to study static as well as dynamics of the excitation, are presented. In Sec.~\ref{sec:Results}, dispersion relations are found using the static Bogoliubov calculation, where roton and maxon modes are identified. We then examine the dynamics of the quantum depletion due to the interaction quench. Excitations of the roton and maxon modes are studied using a self-consistent Bogoliubov method. The asymptotic behavior of the BEC at long times is also explored. Finally, with Sec.~\ref{sec:Conclusion} we conclude our work.

\section{Hamiltonian and Method}\label{sec:Hamiltonian}
\subsection{Hamiltonian of the Rydberg-dressed BEC}
We consider a uniform 3D Bose gas of $N$ atoms that interact through both s-wave and soft-core interactions. The Hamiltonian of the system is given by ($\hbar\equiv 1$),
\BE
\hh \1 \int \psi\dagg({\mathbf{r}})\bigg(-\frac{\nabla^2}{2m}-\mu\bigg)\psi({\mathbf{r}})\dif \mathbf{r}\nn\\
	\2\frac{1}{2}\int\psi\dagg({\mathbf{r}})\psi\dagg(\mathbf{r}')\tilde g(\mathbf{r}-\mathbf{r}')\psi(\mathbf{r})\psi(\mathbf{r}')\dif \mathbf{r}\dif \mathbf{r}',\label{secondquant}
\EE
where $\psi(\mathbf{r})$ is the annihilation operator of the bosonic field,  $\mu$ is the chemical potential, $m$ is the mass of a boson, and $\nabla$ is the 3D nabla operator on coordinate $\mathbf{r}=\{x,y,z\}$. The interaction potential is described by $\tilde g (\mathbf{r}-\mathbf{r}')=g_0\delta(\mathbf{r})+\tilde V(\mathbf{r}-\mathbf{r}')$, where $g_0=4\pi a_{\mathrm s}/m$ is the short-range contact interaction controlled by the s-wave scattering length $a_{\mathrm s}$ \cite{Pethick2008}. $\tilde V(\mathbf{r}-\mathbf{r}')$ is the long-range soft-core interaction, 
\begin{equation}
\tilde V(\mathbf{r}-\mathbf{r}')=\frac{C_0}{R^6+|\mathbf{r}-\mathbf{r}'|^6},
\end{equation}
where $C_0$ is the strength of the dressed interaction potential and  $R$ is the soft-core radius~\cite{Henkel2010a}. Both these parameters can be tuned independently by varying the dressing laser~\cite{Henkel2010a}. The interaction potential saturates to a constant, i.e. $\tilde{V}(\mathbf{r})\rightarrow C_0/R^6$ when $|\mathbf{r}|\ll R$, and approaches to a vdW type at distances of  $|\mathbf{r}| \gg R$, i.e. $\tilde{V}(\mathbf{r})\rightarrow C_0/|\mathbf{r}|^6$  . An example of the soft-core potential is shown in Fig.~\ref{fig:softcore}(a). 
The Fourier transformation of the soft-core potential is
$V(k) = U_{0}f(k)$, where $U_0=C_0/R^6$ determines the strength and $f(k)$ has an analytical form
\BE
f(k) \1 \frac{2\pi^2\ee{-\frac{kR}{2}}}{3kR} \left[\ee{-\frac{kR}{2}}-2\sin\left(\frac{\pi}{6}-\frac{\sqrt{3}kR}{2}\right) \right],\nonumber\label{eq:fk}
\EE
which characterizes the momentum dependence of the interaction. Though the interaction is repulsive in real space, i.e. $\tilde{V}(\mathbf{r})>0$, it contains negative regions in momentum space, as shown in Fig.~\ref{fig:softcore}(b). The negative part of $V(k)$ appears at momentum around  $kR\sim 5\pi/3$. Previously, it was shown that the attractive interaction in momentum space is crucially important to the formation of roton instabilities~\cite{santos_roton-maxon_2003}. 

\subsection{Time-independent Bogoliubov approach}
In momentum space, we expand the field operators using a plane wave basis, $\psi(\mathbf{r})=1/\sqrt{\Omega}\sum_{\mathbf{k}}\ee{\ii {\mathbf{k}}\cdot {\mathbf{r}}}\ah{{\mathbf{k}}}$.  The many-body Hamiltonian can be rewritten as
\begin{equation} \hh=\sum_{{\mathbf{k}}}(\epsilon_{k}-\mu)\ada+\sum_{\mathbf{q},{\mathbf{k}},{\mathbf{k}}'}\frac{g_k}{2\Omega}\hat{a}_{\mathbf{k}+\mathbf{q}}^{\dagger}\hat{a}_{\mathbf{k'}-\mathbf{q}}\dagg\hat{a}_\mathbf{k}\hat{a}_{\mathbf{k'}}, 
\label{Ham_k}
\end{equation}
where $\hat{a}_{\mathbf k}\dagg$ ($\hat{a}_{\mathbf k}$) is the creation (annihilation) operator of the momentum state ${\mathbf{k}}$, and $\Omega$ volume of the BEC. The kinetic energy is $\ek=k^2/2m$ with $k=|\mathbf k|$, while the Fourier transformation of the atomic interaction $\tilde{g}(\mathbf{r}-\mathbf{r}')$ is given by $g_k=g_0+V(k)$. 

For a homogeneous condensate and in the stationary regime, we apply a conventional Bogoliubov approach ~\cite{Bogoliubov1947,YU20082367} to study the excitation spectra. At zero temperature we assume a macroscopic occupation in the condensate, which allows us to replace $\hat{a}_0\approx \sqrt{N_0}$ with $N_0 $ being the number of condensed atoms. We then apply a canonical transformation on the bosonic operators of the non-zero momentum states~\cite{Pethick2008}, $\ah{\mathbf{k}\neq0}=\bar{u}_{k}\bh{\mathbf{k}}-\bar{v}_{k}^*\bh{-\mathbf{k}}\dagg$ where $b_{\mathbf{k}}$ ($\bh{-\mathbf{k}}\dagg$) is the annihilation (creation) operator for bosonic quasiparticles and $\bar{u}_k$ and $\bar{v}_k$ are complex numbers such that $|\bar{u}_k|^2-|\bar{v}_k|^2=1$, which satisfies the bosonic commutation relation~\cite{Pethick2008}. The excitation spectra of the Bogoliubov modes for different momentum components gives the dispersion relation, 
\begin{equation}
\bar{E}_{{k}}=\sqrt{\ek[\ek+2g_kn_0]},\label{Ek}
\end{equation}
with $n_{\mathrm{0}} = N_0/\Omega$ being the density of the condensed atoms.  The coefficients in the Bogoliubov transformation are~\cite{Pethick2008}
\begin{eqnarray}
\bar{u}_k &=&  \sqrt{\frac{1}{2}\left[\frac{\ek+g_kn_0}{\bar{E}_k}+1\right]} \nonumber\\
\bar{v}_k &=& -\sqrt{\frac{1}{2}\left[\frac{\ek+g_kn_0}{\bar{E}_k}-1\right]}
\label{ukvk}.
\end{eqnarray}
The distribution of the non-condensed atoms is given by $n_k = \langle a_k^{\dagger}a_k\rangle = |\bar{v}_k|^2$. Taking into account contributions from all non-condensed components, the quantum depletion in the stationary state is evaluated as $\bar{n}_\text{d}  =1/\Omega \sum_{\textbf{k}\neq 0} |\bar{v}_k|^2$.

\subsection{Self-consistent Bogoliubov approach for the quench dynamics}
\label{timeBogoliubov}
The quench of the soft-core interaction consists of two steps. The system is initially in the ground state of a weakly interacting BEC, i.e. $U_0=0$ when $t<0$. At time $t\geq 0$ the Rydberg dressing is switched on immediately. The scheme is depicted in Fig.~\ref{fig:softcore}(c). The time dependence of the atomic interaction is described by a piecewise function as follows, 
\BE
g_k \1  \left\{
\begin{array}{ll}
	g_0 &  \text{when } t \leq 0 \\
	g_0+U_0f(k) & \text{when } t\geq 0.\\
\end{array} 
\right. 
\EE
We assume that the s-wave interaction is not affected during the quench. Hence we use parameter $\alpha=U_0/g_0$ to characterize the strength of the soft-core interaction with respect to the s-wave interaction. 

A time-dependent Bogoliubov approach is applied to study the dynamics induced by the interaction quench.  It is an extension of the conventional Bogoliubov approximation, where the canonical transformation becomes time-dependent, $\ah{\mathbf{k}\neq0}(t) =u_k(t)\bh{\mathbf{k}}-v_k(t)^*\bh{-\mathbf{k}}\dagg$ where $u_k(t)$ and $v_k(t)$ are time-dependent amplitudes with the relation $|u_k(t)|^2-|v_k(t)|^2=1$, which preserves the bosonic commutation relation. This approach has been widely used to study excitation dynamics in BECs with or without long-range interactions~\cite{Natu2013,Natu2014a,Martone2018,Yin2013a,Kain2014}. It provides a good approximation when the condensate has not undergone significant depletion.

Using the Heisenberg equation of the bosonic operators, we obtain equations of motion of $u_k(t)$ and $v_k(t)$,
\BE
	i\begin{bmatrix}
		\dot{u}_k(t) \\ \dot{v}_k(t)
	\end{bmatrix}= \begin{bmatrix} \tek+g_kn_{\mathrm{c}}(t) & g_kn_{\mathrm{c}}(t) \\ -g_kn_{\mathrm{c}}(t) &-\tek-g_kn_{\mathrm{c}}(t) \end{bmatrix}  \begin{bmatrix}
	{u}_k(t) \\ {v}_k(t)
\end{bmatrix}, \label{Heisenberg}
\EE
where $\tnc$ is the time-dependent condensate density. The total density consists of the condensate and depletion densities as $n = n_{\mathrm{c}}(t) + n_{\rm d}(t)$ with the total density of the excitation, i.e. quantum depletion given as
\begin{equation}
n_{\rm d}(t)=\frac{1}{\Omega}\sum_k n_k(t),
\end{equation} 
where $n_k(t)\equiv\braket{\ada}=|v_{\mathrm{k}}(t)|^2$ is the distribution of all possible momentum states.

For a particle conserving system, both the depleted density and the condensate density are time dependent. In practice, the quantum depletion as a function of time is difficult to calculate, as the differential equations (\ref{Heisenberg}) become non-autonomous. Here we will apply a self-consistent procedure as used in Ref.~\cite{Yin2013a}. First, we force the pre-quench density to be the total density, meaning $n_{\mathrm c}(0)= n$, i.e. assuming that the non-condensed occupation is negligible. This is a valid assumption so long as the s-wave interaction is weak. Eq.~(\ref{Heisenberg}) is solved exactly, yielding solutions
\BE
	 \begin{bmatrix}
		{u}_k(t) \\ {v}_k(t)
	\end{bmatrix}  \1 \bigg[\cos(E_{{k}}(t) t)\mathbb{I} - \mathrm{i}\frac{\sin(E_{{k}}(t) t)}{E_{{k}}(t)} \nn\\	
	& \times &\left(\begin{tabular}{cc} $\tek+g_kn_{\mathrm{c}}(t)$ & $g_kn_{\mathrm{c}}(t)$ \\$-g_kn_{\mathrm{c}}(t)$&$-\tek-g_kn_{\mathrm{c}}(t)$ \end{tabular} \right) \bigg] \begin{bmatrix}
		{u}_k(0) \\ {v}_k(0)
	\end{bmatrix},\nn\\\label{eqnmotion}
\EE
where $\mathbb{I}$ is the identity matrix, and the initial values of $u_k(t)$ and $v_k(t)$ are~\cite{Pethick2008}, 
\begin{eqnarray}
u_k(0) &=&  \sqrt{\frac{1}{2}\left[\frac{\ek+g_0n}{E_k(0)}+1\right]} \nonumber\\
v_k(0) &=& -\sqrt{\frac{1}{2}\left[\frac{\ek+g_0n}{E_k(0)}-1\right]}.
\label{u0v0}
\end{eqnarray}
The evolution of coefficients $u_k(t)$ and $v_k(t)$ depends on the dispersion relation $E_{{k}}(t)=\sqrt{\ek[\ek+2g_kn_{\mathrm{c}}(t)]}$, which is assumed to change adiabatically with time through the condensate density $n_{\rm c}(t)$.
We can then calculate the momentum distribution as
\BE
n_k(t) =& &|v_k(0)|^2+g_k\tnc  \bigg[  g_k\tnc-g_0n\bigg]\nn\\
& \times  & \frac{\tek\left[1-\cos(2E_{{k}}(t) t)\right]}{E_{{k}}(t)^2E_{{k}}(0)}. \label{eq:nk}
\EE
Taking into account all of the momentum components, the quantum depletion is evaluated through,
\BE
n_{\mathrm d}(t) = \frac{1}{2\pi^2}\int_0^{\infty} n_k(t) k^2\dif k \label{depletion},
\EE
where we have replaced the summation by the integration over momentum space. The angular part in the integration has been integrated out in the above equation. With the quantum depletion at hand, the condensate fraction is found to be $n_{\mathrm c}(t)=n-n_{\mathrm d}(t)$. We then reinsert the result back into Eq. (\ref{eq:nk}) and iterate the procedure until the calculation converges self-consistently. 

In the following calculations, we will scale the energies, lengths, and times with respect to the interaction energy $g_0n$, coherence length $\zeta=\left(mg_0n\right)^{-1/2}$, and coherence time $\tau=tg_0n$ of the initial condensate. The zero range  interaction strength is fixed by the s-wave scattering length, which is set to $a_\mathrm{s}=0.1n^{-1/3}$ throughout this work.

\section{Results and discussions}
\label{sec:Results}

\subsection{Stationary dispersion relation}
\begin{figure}
	\includegraphics[width=1.0\linewidth]{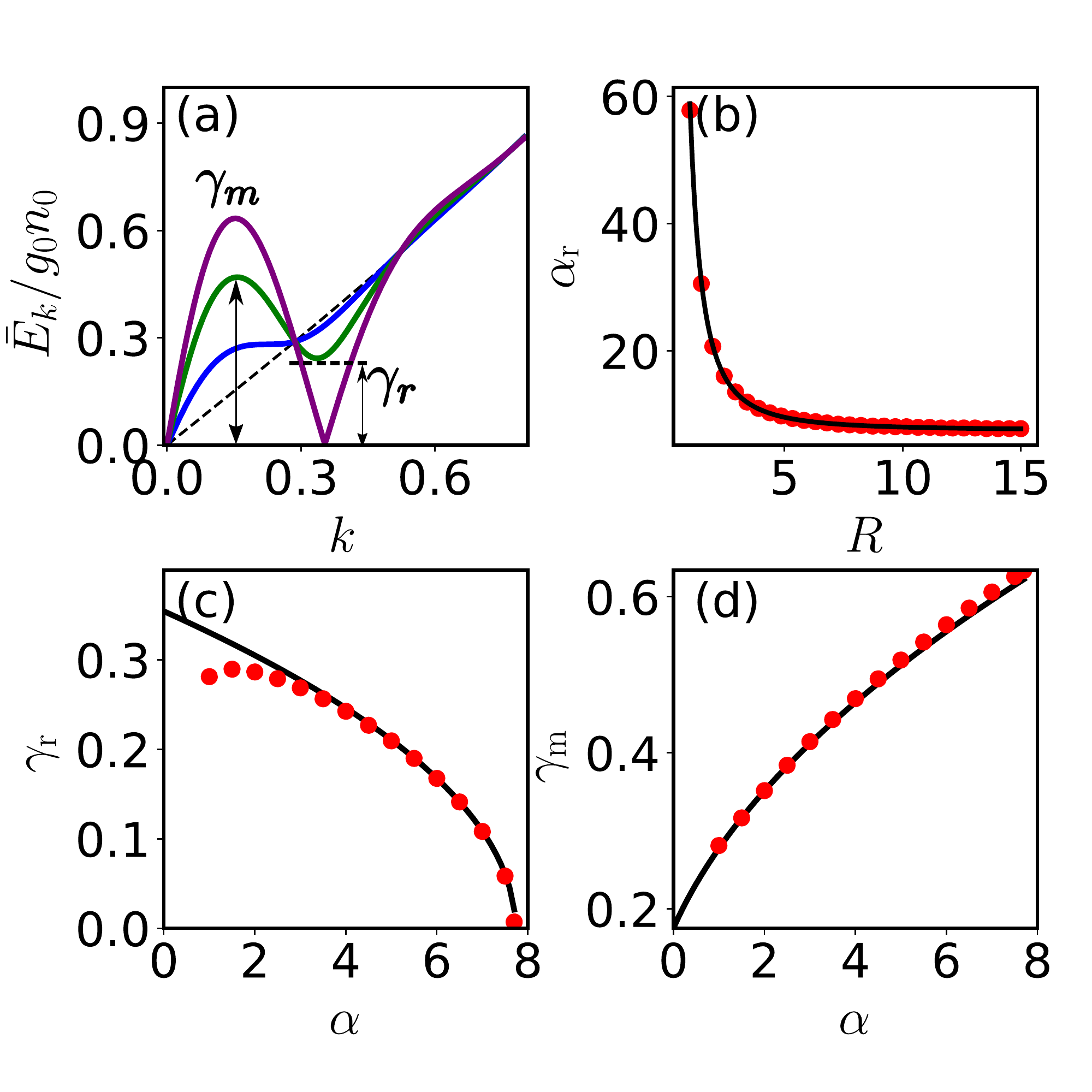}
	\caption{(color online) \textbf{Roton and maxon mode}. (a) Bogoliubov spectra in the stationary state for $\alpha=0$ (dashed), $
1$ (blue), $6$ (green), and $7.7$ (purple), when $R=15$. The energy gaps $\gamma_{\mathrm{r}}$ and $\gamma_{\mathrm{m}}$ indicating respectively the roton and maxon energies are marked for the green curve. For $\alpha>7.7$, the spectra becomes unstable.  (b) The critical value $\alpha_{\mathrm r}$ vs $R$. Analytical calculates (black) agree with the numerical data (red dots). (c) Roton energy $\gamma_{\mathrm{r}}$. Increasing $\alpha$, the roton energy decreases. For large $\alpha$, the analytical (black solid) and numerical (dot) results agree. At small $\alpha$, roton minima become weak and eventually disappear, which leads to the deviation.  The data points in red are the energies taken numerically from the dispersion. (d) Maxon energy $\gamma_{\mathrm{m}}$ increases with $\alpha$. The analytical (black solid) and numerical data agree nicely. In (c) and (d) $R=15$.}	
	\label{fig:dispersion}
\end{figure}

The soft-core interaction drastically alters the dispersion relation of the Bogoliubov excitations. To illustrate this, we first examine dispersion relations of a static BEC by assuming that the soft-core interaction is present. When the soft-core interaction is weak, i.e. $\alpha$ is small, the dispersion relation resembles that of a weakly interacting BEC. The excitation energies increase monotonically with momentum $k$~\cite{Pethick2008} [see Fig.~\ref{fig:dispersion}(a)]. By increasing $\alpha$, the shape of the Bogoliubov spectra changes significantly. A local maximum and minimum can be seen in the dispersion relation [Fig.~\ref{fig:dispersion}(a)]. At the maximum, special modes called maxon modes form, while roton modes emerge around the minima~\cite{santos_roton-maxon_2003}. In the following, we will denote the energies of the maxons and rotons with $\gamma_{\mathrm m}$ and $\gamma_{\mathrm r}$, as the local maximal and minimal values of the dispersion relation. 

The roton and maxon modes depend on the soft-core interaction non-trivially. When increasing $\alpha$, $\gamma_{\mathrm r}$ decreases while $\gamma_{\mathrm m}$ increases, as given by the examples shown in Fig.~\ref{fig:dispersion}(a). For sufficiently large $\alpha$, the roton gap vanishes as the energies become complex, i.e. the roton is unstable. The roton instability can drive the system out of a uniform condensate, leading to the formation of supersolids~\cite{Henkel2010a,Roccuzzo2018a,Macri2013}. It should be noted that the instability here is induced by stronger, isotropic interactions. In dipolar BECs, instabilities are caused by angular dependent interactions \cite{Santos2000}.
\begin{figure*}[t!]
	\includegraphics[width=1.0\linewidth]{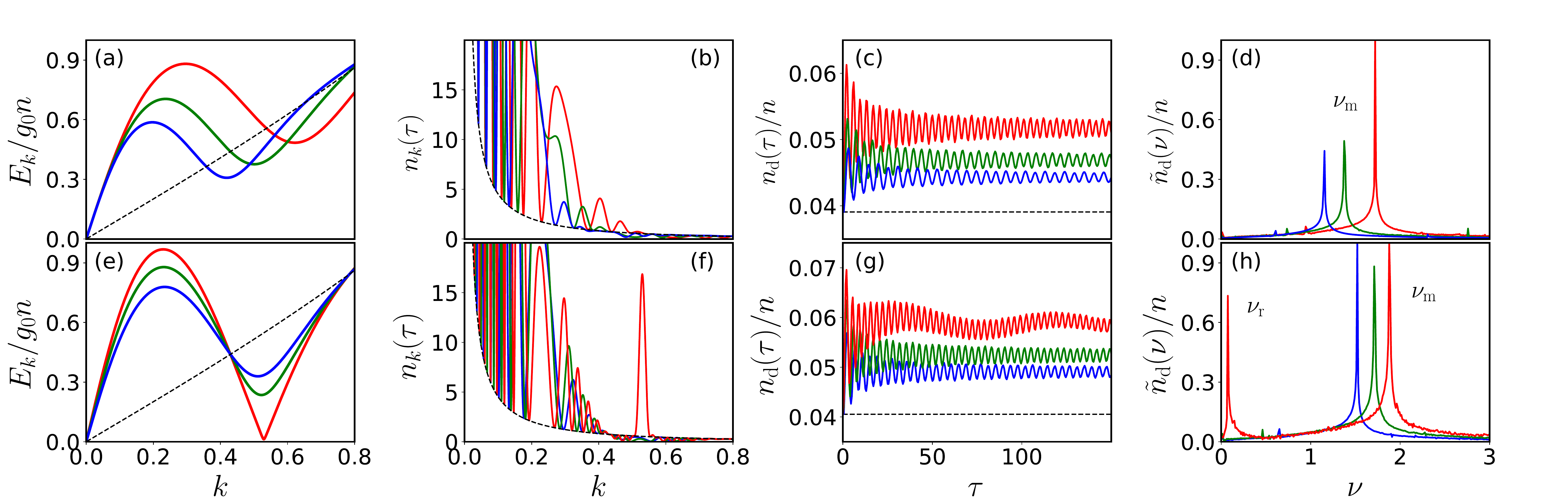}
	\caption{(color online) \textbf{Excitation of the roton and maxon mode}. In the upper panels, (a) gives the dispersion for a static BEC. The momentum of the roton and maxon modes decreases with increasing soft-core radius $R$. Without soft-core interactions, the excitation energy monotonically increases with momentum (dashed). In (b)-(c), the interaction quench is applied. Momentum densities at time $\tau=30$ are shown in (b).  The black dashed curve shows the momentum distribution of the initial state. Fast oscillations are found in the quantum depletion (c), which leads to sharp peaks in the respective Fourier transformation (d). The frequency $\mu_{\rm m}$ at the major peaks is determined by the maxon frequency. Minor peaks corresponding to other frequencies are almost invisible. In (a)-(d), three different soft-core radius $R=8$ (red), $10$ (green), and $12$ (blue) are considered. In the lower panels,  the dispersion (e), momentum distribution (f), quantum depletion (g) and Fourier transformation of the quantum depletion (h) for $R=10$ and $\alpha=5$ (blue), $6.5$ (green), and $7.99$ (red) are shown. Approaching to the roton instability (e), the momentum distribution (f) develops a large occupation around modes at $k_{\mathrm r}$ at $\tau=30$. The depletion dynamics maintains a slower oscillation (g) as the interaction strength is increased, which can be seen from the Fourier transformation of the quantum depltion (h). The lower peak frequency $\nu_{\rm r}$ is determined by the roton mode. The major peaks at higher frequencies are due to the excitation of maxons. When $\alpha = 7.99$, both the roton and maxon mode are dynamically stable, giving narrow Fourier spectra.}
	\label{fig:BECdensity}
\end{figure*}

It is important to obtain the critical value at which the roton mode becomes unstable. From Fig.~\ref{fig:softcore}(b), the Fourier transform of the soft-core potential has the most negative value around $k_{\mathrm r} \approx 5\pi/3R$. The roton minimum takes place around this momentum. By substituting $k_{\mathrm{r}}$ into the dispersion relation, we can identify the critical $\alpha$ at which the roton energy becomes complex, 
\begin{eqnarray}
\alpha_\mathrm{r} &=& \frac{5 \ee{5 \pi /3} \left(36  R^2+25 \pi ^2\right)}{72 \pi  R^2 \left[2 \ee{5 \pi /6} \sin \left(\frac{\pi }{6}-\frac{5 \pi }{2 \sqrt{3}}\right)-1\right]} .
\end{eqnarray}  
To check the accuracy of this critical value, we numerically find the instability point from the dispersion relation. Both numerical and analytical values are shown in Fig.~\ref{fig:dispersion}(b). The analytical result agrees with the numerical values very well. This supports the assumption that the roton minimum happens around momentum $k_{\mathrm r}$. 

Knowing the momentum $k_{\rm r}$, we can obtain the roton energies by inserting it into Eq. (\ref{Ek}). It is found that the roton energy $\gamma_{\mathrm r}$ decreases with increasing $\alpha$ [see Fig. ~\ref{fig:dispersion}(c)]. The roton energy from the numerical calculations agrees with the analytical data, especially when the soft-core interaction is strong. Decreasing the soft-core interaction,  the roton modes disappear for sufficiently small $\alpha$, as our numerical calculations indicate. We notice large deviations between the two methods in this regime. 

On the other hand, the location of the maxon modes in momentum space is difficult to find. By analyzing the dispersion relation, the momentum corresponding to the maxon mode is approximately given by $k_{\mathrm m}\approx k_{\mathrm r}/2$. Using this approximation, we substitute this momentum value into Eq. (\ref{Ek}) and calculate the maxon energy. The result is shown in Fig. \ref{fig:dispersion}(d), where the approximate value matches the numerical values with a high degree of accuracy.

Recently, the stationary state of 2D and 3D Rydberg-dressed BECs have been examined \cite{Seydi2020}. It was shown that the increased occupation around the roton modes leads to instabilities in the ground state in the form of density waves. It was also seen that the strong interparticle interactions lead to a large depletion of the condensate.

\subsection{Roton and maxon excitation}
\label{subsec:Quantum Depletion}
 Depending on parameters of the soft-core interaction, the stationary dispersion relation could support roton and maxon modes. One example is displayed in Fig. \ref{fig:BECdensity}(a). Now if we quench the interaction, the dispersion relation of the initial and final state is different. The system is driven out of equilibrium, such that momentum distributions evolve with time. In Fig \ref{fig:BECdensity}(b), snapshots of the momentum distribution are shown. At $\tau=0$, the BEC is in a stationary state, which depends on the initial condition, $\bar{v}_k$. The respective momentum distribution is a smooth function of $k$. At later times, different momentum components are excited by the presence of the soft-core interaction, causing dynamical evolution of the quantum depletion.  

The dynamics of the quantum depletion depends vitally on the parameter $R$ and $\alpha$ in the soft-core interaction. After switching on the interaction, the excitation of the Bogoliubov modes significantly affects the momentum distribution. We will first investigate the oscillatory behavior of the quantum depletion. For moderate soft-core interactions, many momentum modes are excited by the soft-core interaction, as shown in Fig.~\ref{fig:BECdensity}(c). As a result, the quantum depletion increases rapidly with time, and then oscillates around a constant value [Fig.~\ref{fig:BECdensity}(c)]. The Fourier transformation $\tilde{n}_{\mathrm c}(\nu)$ of the quantum depletion, characterizing the spectra of the dynamics, shows a sharp peak [Fig.~\ref{fig:BECdensity}(d)]. The peak positions, i.e. frequency of the oscillations, decrease gradually when increasing the soft-core radius. 

For stronger soft-core interactions, the roton mode moves towards the instability point [see Fig.~\ref{fig:BECdensity}(e)]. In this case, higher momentum components can be excited during the interaction quench [Fig.~\ref{fig:BECdensity}(f)]. Here a new, lower frequency pattern develops on top of the fast oscillation in the quantum depletion [Fig.~\ref{fig:BECdensity}(g)]. This changes the Fourier spectra of the quantum depletion, where a new peak is found at a lower frequency [Fig.~\ref{fig:BECdensity}(h)]. 

\begin{figure}
\centering
\includegraphics[width=1.0\linewidth]{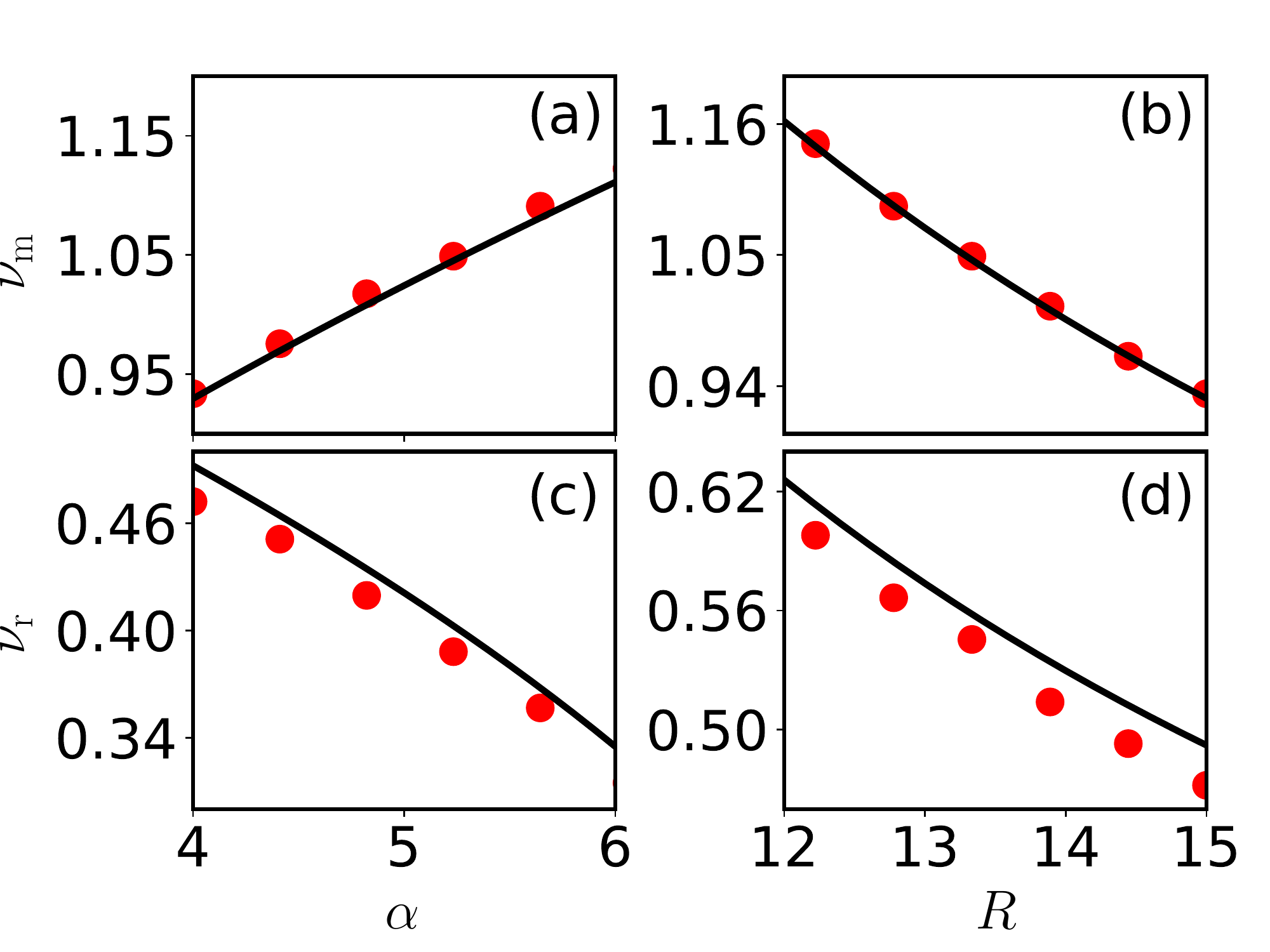}
\caption{(color online) \textbf{Maxon frequency (a-b) and roton frequency (c-d)}. The dots are numerical data from the Fourier spectra. The solid curves are analytical results $2\gamma_{\rm m}$ in (a)-(b) and $2\gamma_{\rm r}$ in (c)-(d) obtained from the Bogoliubov dispersion. The maxon (roton) frequency increases (decreases) with increasing interaction strength. At the critical point $\alpha_{\rm r}$, the roton mode loses stability. Frequencies of both modes tends towards $0$ for larger $R$ values as the soft-core interaction becomes weaker. In (a) and (c) $R=15$. In (b) and (d) $\alpha=4$. }
\label{fig:frequencies}
\end{figure}

It is important that the peak positions in $\tilde{n}_{\mathrm c}(\nu)$ are determined by the roton and maxon energies. In the quantum depletion, the fast oscillations are due to the excitations of the maxon modes, while slow oscillations are due to the roton modes. To verify this, we first obtain the maxon and roton frequencies by substituting the corresponding momentum $k_{\rm m}$ and $k_{\rm r}$ in Eq.~(\ref{Ek}). We then compare them with the frequency at the peak positions in the Fourier spectra. Note that the oscillation frequency (i.e. peak frequency of the Fourier spectra) in the quantum depletion is twice the Bogoliubov energy, as can be seen in Eq.~(\ref{eq:nk}). As shown in Fig.~\ref{fig:frequencies}, the numerical data for both the maxon mode (a-b) and roton mode (c-d) agree with the analytical calculations. When varying the interaction strength, the maxon (roton) frequency increases (decreases) with increasing $\alpha$. If we increase the soft-core radius, frequencies of both modes decrease. 

The agreement between numerical and analytical calculation confirm that both roton and maxon modes are excited via quenching the soft-core interaction. The dynamically excited modes are stable, as both the fast and slow oscillations are {\it persistent} for a long time. In our numerical simulations, the oscillations will not dampen even when the simulation time $\tau>1000$. Such persistent oscillatory dynamics also leads to the sharp peaks in the Fourier transformation of the quantum depletion. 

We want to emphasize that the quench dynamics in the dressed BEC is in sharp contrast to BECs with either s-wave or dipolar interactions. In a weakly interacting BEC, the quantum depletion grows exponentially to a steady value $\propto\zeta^{-\frac{1}{3}}$, while oscillatory patterns are not present in the depletion~\cite{Natu2013}, due to the fact that low energy phonon modes dominate the quench dynamics. In dipolar BECs~\cite{Tian2018,Chomaz2018,Natu2014a,Griesmaier2005,Lahaye2009}, on the other hand, roton modes are formed due to the interplay between long-range dipolar and s-wave interactions~\cite{Tian2018,Chomaz2018,Natu2014a,Griesmaier2005,Lahaye2009}. These roton modes can be excited by quenching the dipolar interaction, while maxon modes are typically unstable in the dynamics [see Appendix~\ref{app:dipolar} for examples].

\begin{figure}[t]
	\includegraphics[width=1.0\linewidth]{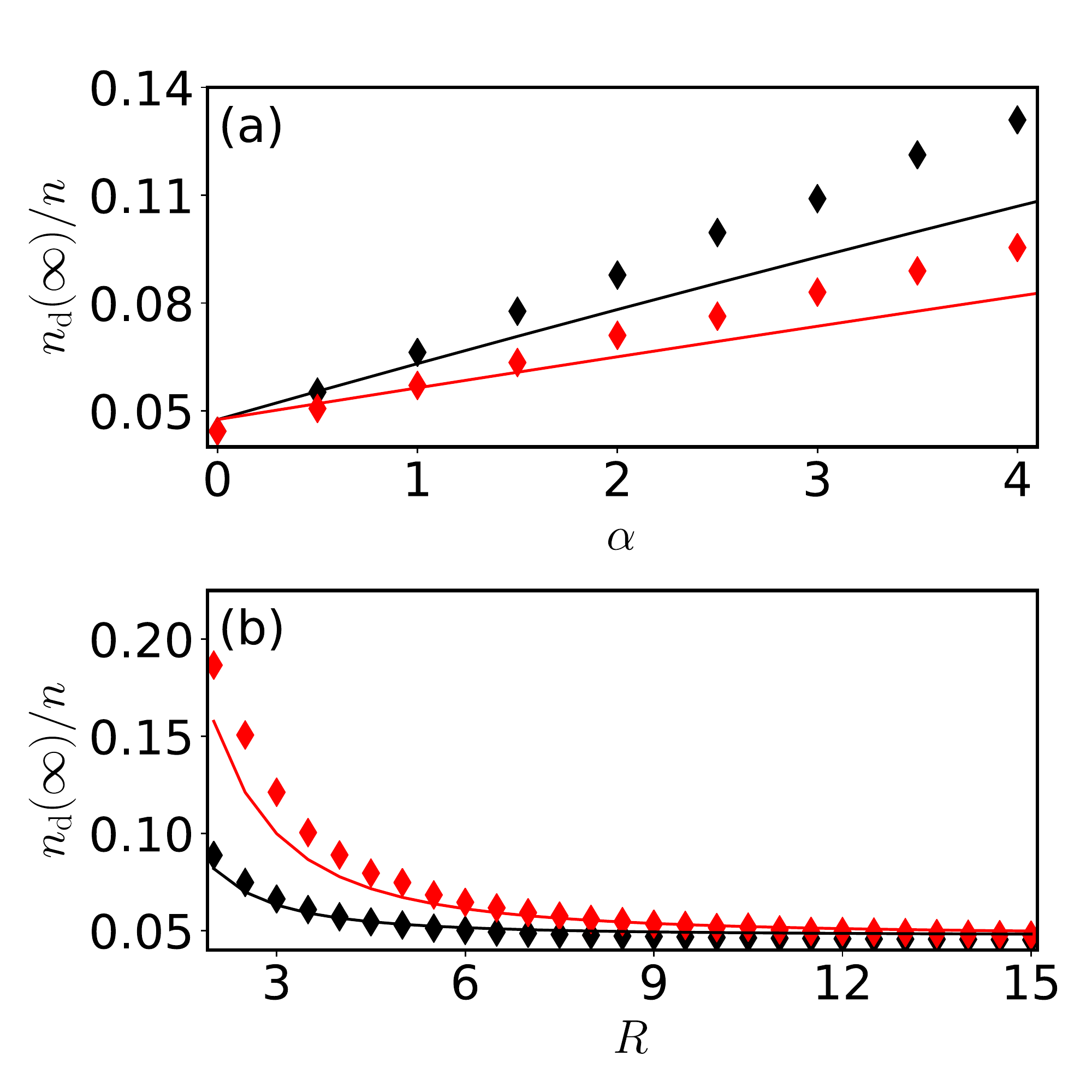}
	\caption{(color online) \textbf{Asymptotic quantum depletion}. The asymptotic quantum depletion increases with increasing $\alpha$ (a), which is seen from both the analytical and numerical calculations. The quantum depletion $ n_{\textrm{d}}^{\infty}$ decreases with increasing soft-core radius (b). The solid line is found analytically using Eq. (\ref{eq:ndinf}), while the data points are found by numerically solving Eq. (\ref{depletion}) and taking the mean value at later times between $\tau\approx50\rightarrow150$. Parameters in (a) are $R=3$ (black) and $4$ (red). Parameters in (b) are $\alpha=1$ (black) and $3.5$ (red). }
	\label{fig:asymptotic}
\end{figure}
\subsection{Quantum depletion in the long time limit}
In the long time limit $\tau \gg 1$, the quantum depletion oscillates rapidly around a mean value [Fig.~\ref{fig:BECdensity}(c) and (g)]. In the following, we will evaluate the asymptotic value of the quantum depletion. First we will derive an analytic expression using the following approximations. In the long time limit, the time averaged quantum depletion is largely determined by the low momentum modes. Moreover, we will neglect the oscillation term in Eq.~(\ref{eq:nk}), as they are related to the roton and maxons. Using these approximations, the asymptotic form of the momentum distribution $n_k^{\infty}$ is obtained,
\begin{equation}
n_k^{\infty}\approx 	\frac{1}{2} \left(   \frac{k^2+1}{\sqrt{k^2(k^2+4)}} -1 \right)		  + \frac{\alpha f(k)}{4 k}\frac{n_{\mathrm{c}}^{\infty}}{n},
\label{eq:nkinf}
\end{equation}
where $n_{\rm c}^{{\infty}}$ is the asymptotic condensate density.  After carrying out the integral over  momentum space,  the approximate quantum depletion when $\tau\to \infty$ is obtained,
\begin{equation}
\frac{n_{\textrm{d}}^{\infty}}{n} \approx 2\Gamma\left(\frac{R^2+\alpha\pi}{3R^2+2\pi\alpha\Gamma}\right),
\label{eq:ndinf}
\end{equation}
where $\Gamma=(2\pi^2\zeta^3n)^{-1}$. This result predicts that the quantum depletion approaches to a constant value $n_{\rm d}^{\infty}/n\to 2\Gamma/3$ in the limit $R\to \infty$. This resembles the result of the weakly interacting BEC, i.e. the soft-core interactions plays no role in the quench. 

To verify the analytical calculation, we numerically find the mean value of the quantum depletion when time is large. Both the numerical and analytical results are shown in Fig.~\ref{fig:asymptotic}.  For small $\alpha$, low momentum states are populated by switching on the soft-core interaction. This is the regime where the approximation works. We find a good agreement between the numerical and analytical calculations. Increasing the interaction strength, more and more higher momentum components are populated [see Fig.~\ref{fig:BECdensity}(b) and (f)], causing larger depletion. A clear deviation between the numerical and analytical data is found, as the approximation we made in evaluating Eq.~(\ref{eq:ndinf}) becomes less accurate. On the other hand, the quantum depletion becomes smaller by increasing the soft-core radius, as the strength of the soft-core interaction reduces. In this case the numerical and analytical results agree well [see Fig.~\ref{fig:asymptotic}(b)].

\section{Conclusion}\label{sec:Conclusion}
We have studied dynamics of 3D BECs in free space, with Rydberg-dressed soft-core interactions. An interaction quench is implemented through turning on the soft-core interaction instantaneously, starting from a weakly interacting BEC. The Bogoliubov spectra of the BEC displays local maxima and minima, which are identified as maxon and roton modes. Through a time-dependent Bogoliubov approach, we have calculated dynamics of the quantum depletion self-consistently. Our results show that both roton and maxon modes are excited by switching on the soft-core interaction. The excitation of roton and maxon modes generate slow and fast oscillatory dynamics in the quantum depletion. Our simulations show that the excited roton and maxon mode are stable in the presence of the soft-core interaction, which are observed from the persistent oscillations of the quantum depletion. We have found the frequencies of the roton and maxon modes approximately, which are confirmed by the numerical simulations. 
 
Our study shows that exotic roton and maxon excitations can be created in Rydberg-dressed BECs through the interaction quench. Properties of the maxons and rotons can also been seen from condensate fluctuations [see Appendix~\ref{sec:number} for details)] and density-density correlations [see Appendix \ref{sec:correlation}]. The result studied in this work might be useful to identify the soft-core interaction through measuring frequencies and strength of the quantum depletion. In the future, it is worth studying formation of droplets and spatial patterns in Rydberg-dressed BECs, which could be affected by the presence of roton or maxon modes. 

\section*{Acknowledgements}
We thank Yijia Zhou and S Kumar Mallavarapu for fruitful discussions. The research leading to these results has received funding from the EPSRC Grant No. EP/M014266/1, the EPSRC Grant No. EP/R04340X/1 via the QuantERA project “ERyQSenS”, the UKIERI-UGC Thematic Partnership No. IND/CONT/G/16-17/73, and the Royal Society through the International Exchanges Cost Share award No. IEC$\backslash$NSFC$\backslash$181078. We are grateful for access to the Augusta High Performance Computing Facility in the University of Nottingham.
 
\begin{appendix}
\renewcommand{\thefigure}{A\arabic{figure}}
\setcounter{figure}{0}

\section{Dynamics of 2D Dipolar Systems}\label{app:dipolar}
Quench dynamics in BECs with dipolar interactions are drastically different. The dipolar interaction is given by 
\BE
\tilde{V}_{\text{dd}}(\mathbf{r}-\bold{r}')\1 g_0\delta(\mathbf{r}) + \frac{d^2}{|\mathbf{r}-\bold{r}'|^3}[1-3\cos^2(\theta)],
\label{se:dipolar}
\EE
 where $d$ is the dipole moment, $\theta$ is the angle between the dipoles and molecular axis, and $g_0$ is the short-range contact interaction as before. In 3D, the Fourier transform of the dipolar interaction has no momentum dependence \cite{Lahaye_2009}. In a 2D trapped dipolar Bose gas~\cite{Fischer2006a,Ticknor2011a}, the interaction potential displays a strong momentum dependence~\cite{Natu2014a}.

We consider a quasi-2D setup~\cite{Natu2014a}, where a strong confinement is applied in the perpendicular $z$-direction while leaving atoms free to move in the $x-y$ plane. The dipoles are  polarized along this $z$-axis. This leads the axial confinement as $l_z$, which provides a natural rescaling of $\bold r\mapsto\bold r/ l_z$ \cite{Fischer2006a,Wilson2016,Cai2010,Ticknor2011a,Natu2014a}. After integrating Eq. (\ref{se:dipolar}) in the $z$-axis, we obtain the Fourier transformation of the quasi-2D dipolar interaction~\cite{Natu2014a}
\BE
V_{\mathrm{dd}}(k) =2-3k\sqrt{\pi} \text{Erfc}(k)\ee{k^2},
\EE
where Erfc$(k)$ is the complimentary error function. Here we define the dimensionless parameter $\alpha_{\mathrm d}=d^2/g_0$ to characterizing the strength of the dipolar interaction, such that the interaction after the quench is given as $g_d/g_0=1+\alpha_{\mathrm d} V_{\mathrm{dd}}(k)$.
The quench scheme for the dipolar case is similar to the procedure outlined in the main text.  We switch on the dipolar interaction instantaneously, while keeping the s-wave interaction unchanged. 

\begin{figure}	
	\includegraphics[width=1.0\linewidth]{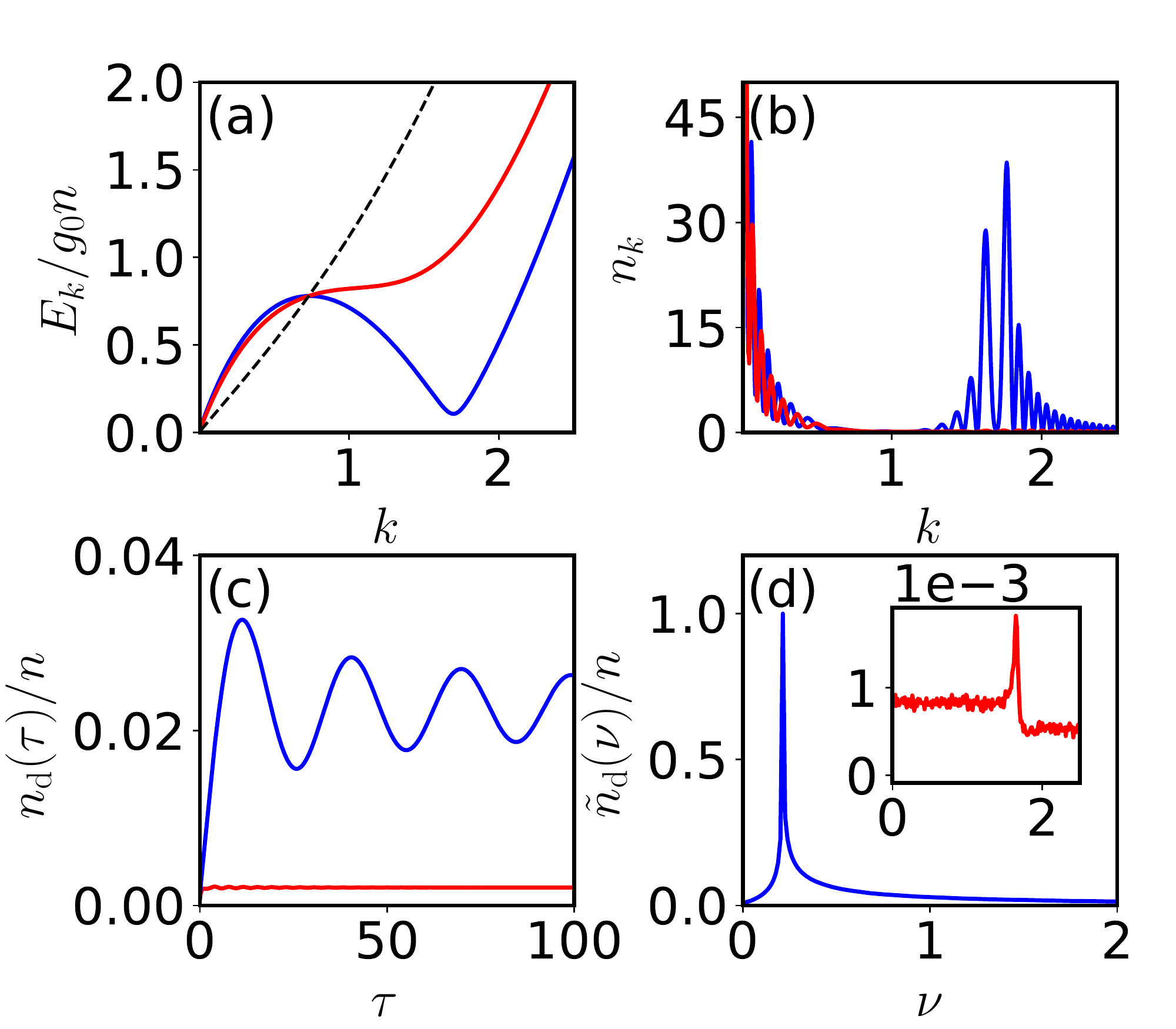}
	\caption{(color online) \textbf{Quantum depletion in a dipolar BEC}. Red curves are for $\alpha_{\rm d}=2.1$ and blue curves are for $\alpha_{\rm d}=2.7$. The axial confinement is set to $l_z=0.1n^{-1/2}$. We show the dispersion relation in (a) while the momentum distribution at time $\tau=30$ is shown for (b). The quantum depletion and corresponding Fourier spectra are shown in (c) and (d) respectively. The inset shows a maxon mode is excited for $\alpha_{\rm d}=2.1$. However the signal is very weak and almost invisible. The axes of the inset is same as panel (d).}
	\label{sfig:dispersion}
\end{figure}

The dispersion relation for the dipolar BEC is shown in Fig. \ref{sfig:dispersion}(a), where both roton and maxon modes can be seen. 

When the dipolar interaction is compared to the Rydberg-dressed BEC [e.g Fig. \ref{fig:dispersion}(a)], the energies of the low momentum modes remain small, as seen by directly comparing the dispersion relations. The absence of these large maxon energies means that the mechanism behind the dipolar interactions prevent the oscillations that we previously attributed to the maxon modes from reaching large amplitudes [Fig. \ref{sfig:dispersion}(b)]\cite{Natu2014a,Wilson2016,Katz2002}.

We follow the same self-consistent process to obtain the condensate fraction. We calculate the quantum depletion as before as
$n_{\rm d}/n= 1/(2\pi l_z^2n)\int_0^{\infty}n_{k}k\dif k$.
When $\alpha_{\rm d}$ is small, the dynamics develops maxon oscillations,  which dampens in short time scales, as shown in Fig. \ref{sfig:dispersion}(c). When $\alpha_{\rm d}$ is large, the roton frequency completely overpowers the maxon frequency in the dynamics. The absence of a stable maxon mode is also seen in the Fourier spectra [Fig. \ref{sfig:dispersion}(d)]. 
\renewcommand{\thefigure}{B\arabic{figure}}
\setcounter{figure}{0}

\section{Condensate fluctuation}\label{sec:number}
\begin{figure}[t]
	\includegraphics[width=0.95\linewidth]{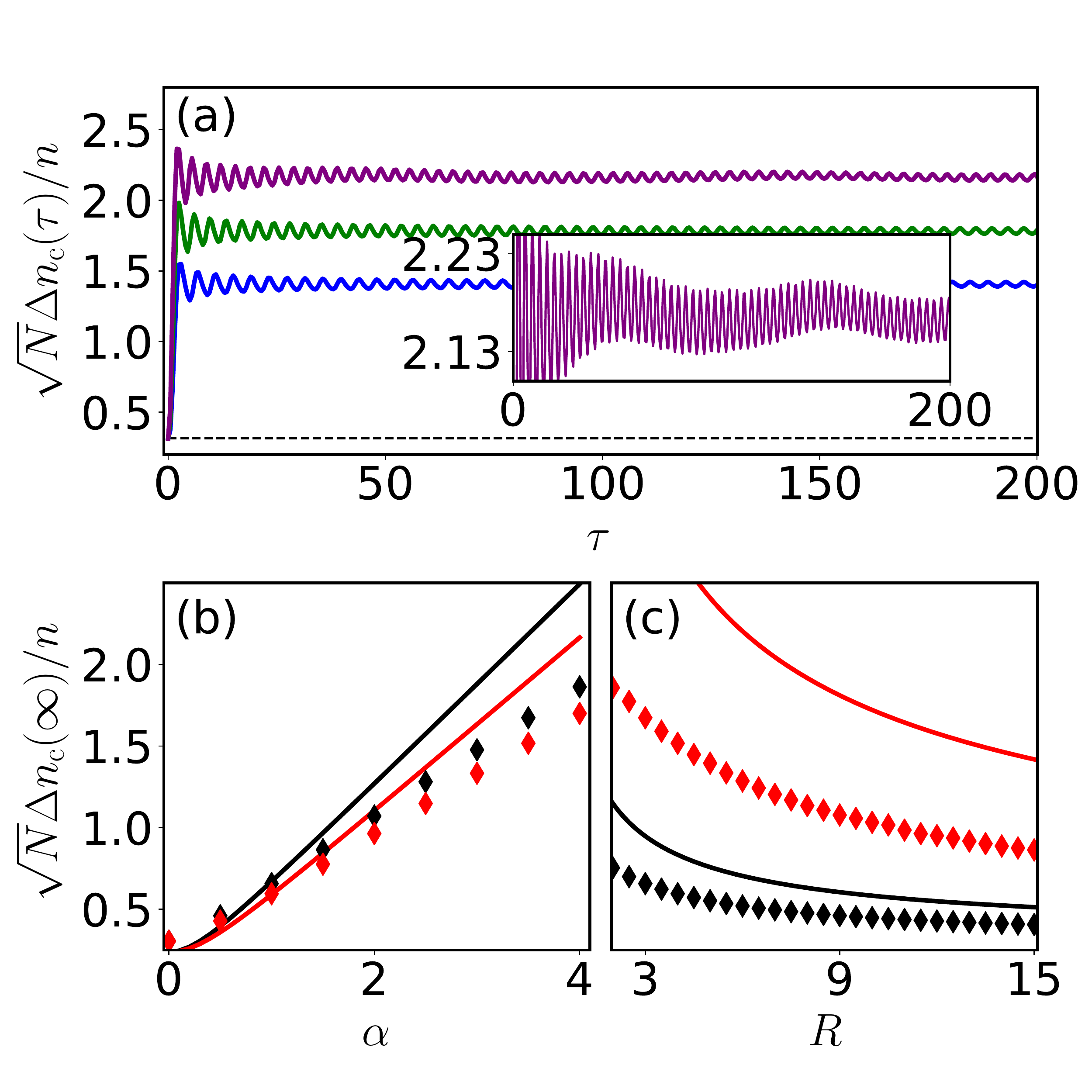}
	\caption{(color online) \textbf{Condensate fluctuation}. (a) Dynamics of the condensate fluctuation. We fix $R=10$, and evolve the system for $\alpha=5$ (blue), $6.5$ (green), and $7.7$ (purple). The dashed line is the fluctuation without the soft-core interaction, i.e. $\alpha=0$. The inset shows fluctuations when $\alpha=7.99$ to highlight the low frequency oscillations due to rotons. Mean values of the fluctuations for different $\alpha$ (b) and $R$ (c) when time $\tau \rightarrow\infty$. We have considered $R=3$ (black) and $4$ (red) in (b) and $\alpha=1$ (black) and $3.5$ (red) in (c). Other parameters can be found in Fig. \ref{fig:asymptotic} in the main text.}\label{sfig:numfluc}
\end{figure}

In this section, we evaluate the fluctuation of the condensate for the Rydberg-dressed BEC. The condensate fluctuation is defined as
\begin{eqnarray}
\Delta n_{\mathrm c} &=& \sqrt{\braket{n_{\mathrm c}^2}-\braket{n_{\mathrm c}}^2} \nonumber\\
&=& \sqrt{\braket{n_{\mathrm d}^2}-\braket{n_{\mathrm d}}^2}\nonumber \\
&=&   \frac{1}{\Omega}\sqrt{\sum_{\bold k\bold k'\neq0}\left[\braket{\hat{a}_{\bold k}\dagg\hat{a}_{\bold k}\hat{a}_{\bold k'}\dagg\hat{a}_{\bold k'}}-\braket{\hat{a}_{\bold k}\dagg\hat{a}_{\bold k}}\braket{\hat{a}_{\bold k'}\dagg\hat{a}_{\bold k'}}\right] }\nn,
\end{eqnarray}
where we have assumed the total density $n$ is a constant. Using the Bogoliubov transformation, the fluctuation of the condensate is obtained,
\BE
\Delta \nc 
	\1 \frac{1}{\Omega}\sqrt{2\sum_{\bold k\neq0}n_k(1+n_k)} \label{se:fluc}
\EE
One can numerically evaluate the fluctuation by inserting Eq.~(\ref{eq:nk}) into the above equation. For convenience, the relative fluctuation, $\sqrt{N}\Delta n_{\rm c}/n$, will be calculated. Some examples are shown in Fig.~\ref{sfig:numfluc}(a). The fluctuation increases rapidly, and then saturates at an asymptotic value when time is large. The fluctuation oscillates around the asymptotic value. The maxon modes lead to fast oscillations. When the roton mode is significantly populated, a slower oscillation is found.

The asymptotic value of the fluctuation depends on  the soft-core interaction. Increasing $\alpha$, the asymptotic value increases [see Fig.~\ref{sfig:numfluc}(a) and (b)]. We can estimate the asymptotic value of the density fluctuation by replacing $n_k$ with its asymptotic value Eq.~(\ref{eq:nkinf}), in Eq.~(\ref{se:fluc}), which yields
\BE
\frac{\sqrt{N}\Delta n_{\mathrm{c}}^{\infty}}{n}\1 \sqrt{2\Gamma\int_0^{\infty}n_k^{\infty}\left[1+n_k^{\infty}\right] k^2 \dif k}. \label{seeq:flucasymp}
\EE
Further assuming the fluctuation depends solely on low momentum states, we obtain the approximate result of the fluctuation when $\tau\to \infty$,
\BE
\frac{\sqrt{N}\Delta n_{\mathrm c}^{\infty}}{n} \approx \sqrt{   \frac{2\Gamma\pi^2\left[  1+\pi^2\alpha\left(  6\sqrt{3} +\pi\alpha C  \right) \right]}{27R}  },
 \label{eqn:analytic_numfluc}
\EE
with the constant $ C= \left[  4\sqrt{3}\pi -3 \log\left(\frac{27}{16}\right)  \right]$. The approximation result shows that fluctuations of the condensate decreases (increases) with increasing $R$ ($\alpha$). In Fig.~\ref{sfig:numfluc}(b) and (c), numerical and approximate results are both shown. The two calculations agree when $\alpha$ is small or $R$ is large, where the depletion and fluctuation are both small. Though large discrepancy is found when $\alpha$ is large or $R$ is small, the trend found from both numerical and analytical calculations are the same.

\renewcommand{\thefigure}{C\arabic{figure}}
\setcounter{figure}{0}

\begin{figure}[t]
	\includegraphics[width=0.9\linewidth]{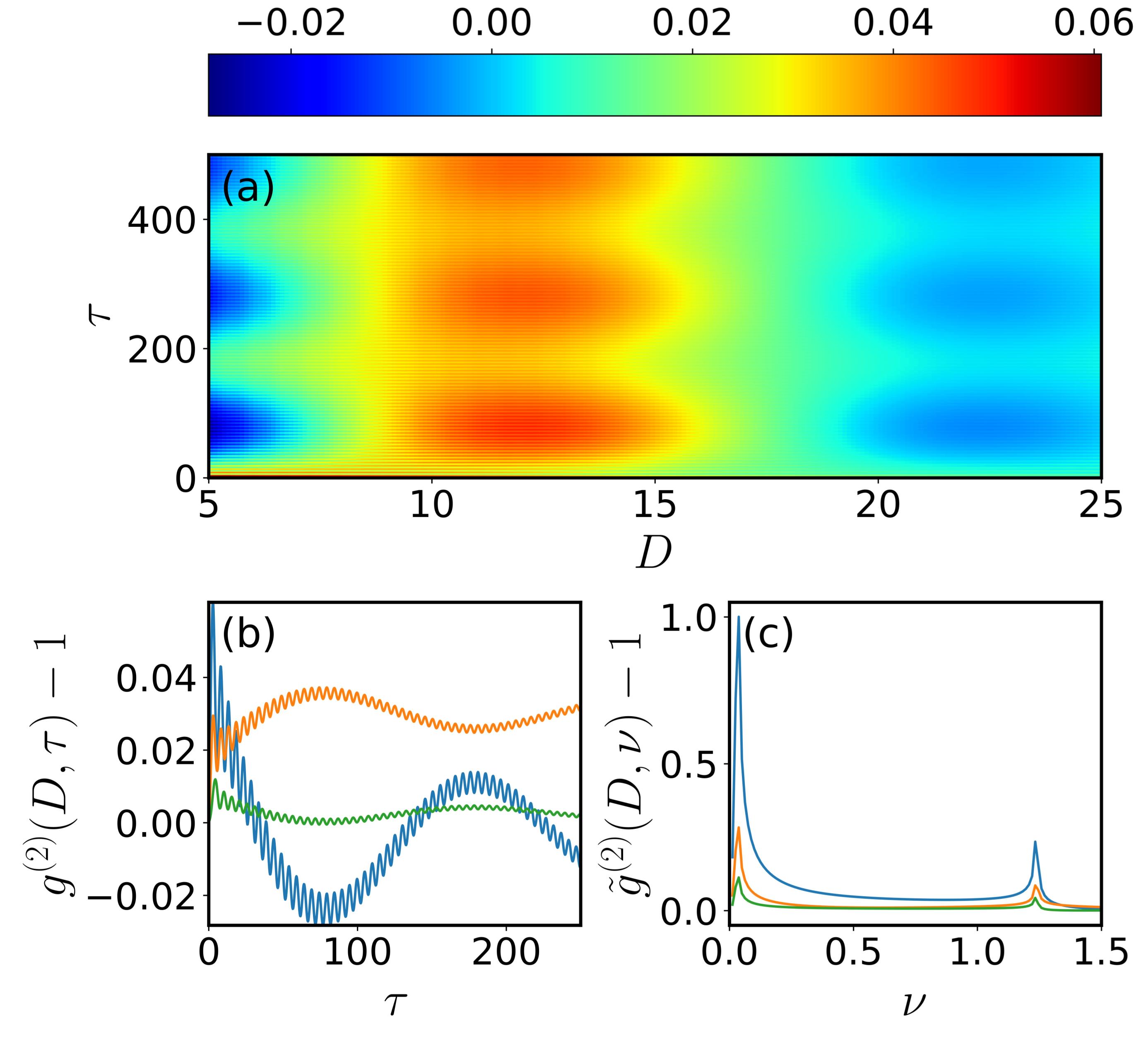}
	\caption{(color online) \textbf{Density-density correlation}. (a) the density-density correlations as a function of $D$ and $\tau$, when $R=15$ and $\alpha=7.7$. Correlations at  $D=5$ (blue), $15$ (orange), and $25$ (green) are shown in (b). The corresponding Fourier spectrum of the correlation function is shown in (c). In the Fourier spectra, the peaks at lower and higher frequencies are due to the excitation of roton and maxon modes.}\label{sfig:corr}
\end{figure}

\section{Density-Density Correlation}\label{sec:correlation}
Lastly we evaluate the density-density correlation function~\cite{Natu2013,Martone2018}
\BE
\gt(\mathbf{r},t)\1 \sum_{\mathbf{k},\mathbf{k'},\mathbf{q}}e^{i \mathbf{k}.\mathbf{r}}\frac{1}{\Omega^2}\left\langle\ah{\mathbf{k}+\mathbf{q}}\dagg(t)\ah{\mathbf{k}}(t)\ah{\mathbf{k'}-\mathbf{q}}\dagg(t)\ah{\mathbf{k'}}(t)\right\rangle\nn\\.
\EE
Within the Bogoliubov transformation, this can then be expressed in terms of the condensate density as $\braket{1/\Omega^2\sum_{\mathbf{k},\mathbf{k'}}\ah{\mathbf{k}+\mathbf{q}}\dagg(t)\ah{\mathbf k}(t)\ah{\mathbf{k'}-\mathbf q}\dagg(t)\ah{\mathbf{k'}}(t)}=n^2+n/\Omega \sum_{\mathbf k}[4\vtwo-u_{\mathbf{k}}^*v_{\mathbf{k}}-u_{\mathbf{k}} v_{\mathbf{k}}^*]$. Defining $D=|\mathbf{r}-\mathbf{r'}\,|/\zeta$ as the scaled interatomic distance, the correlation function is given as \cite{Natu2013}
\BE
\tgt(D,\tau) -1 = \frac{4\Gamma}{D}\int^{\infty}_0 k\dif k\sin(kD)\big[n_k-\text{Re}[u_{k}^*(\tau)v_{k}(\tau)]\big].\nn\\\label{correlation}
\EE

We see from Fig. \ref{sfig:corr}(a) that the correlations immediately develop both slow and fast oscillations. The slow oscillation corresponds to the excitation of roton modes, where $\gamma_{\mathrm r}$ is small. The fast oscillations attributed to the maxon occupation are more easily observed when looking at a specific value of $D$ [Fig. \ref{sfig:corr}(b)]. The corresponding Fourier transformation $\tilde{g}^{(2)}\left(D,\nu \right)-1$ clearly show the associated frequency peaks. When the distance $D<R$, $g^{(2)}(D,\tau)$ oscillates with large amplitudes and can have negative values, i.e. strong repulsive interactions lead to anti-correlations. Around the soft-core radius, the correlations are positive, and reach their maximal values. When $D\gg R$, the correlations tend to zero at large times.

\end{appendix}
 \bibliography{Quenched_Dynamics_Paper}
\end{document}